\documentclass[twocolumn,printnumbers,amsmath,amssymb,prl,verbatim]{revtex4}
\usepackage{graphicx}
\usepackage{subfigure}
\usepackage{color}

\begin{document}

\title{Role of disorder in determining the vibrational properties of mass-spring networks}

\date{\today}

\author{Yunhuan Nie$^1$}
\author{Hua Tong$^{1,2}$}
\author{Jun Liu$^1$}
\author{Mengjie Zu$^1$}
\author{Ning Xu$^{1,*}$}

\affiliation{$^1$CAS Key Laboratory of Soft Matter Chemistry, Hefei National Laboratory for Physical Sciences at the Microscale and Department of Physics, University of Science and Technology of China, Hefei 230026, People's Republic of China\\
$^2$Institute of Industrial Science, University of Tokyo, Meguro-ku, Tokyo 153-8505, Japan}

\begin{abstract}

By introducing four fundamental types of disorders into a two-dimensional triangular lattice separately, we determine the role of each type of disorder in the vibration of the resulting mass-spring networks. We are concerned mainly with the origin of the boson peak and the connection between the boson peak and the transverse Ioffe--Regel limit. For all types of disorders, we observe the emergence of the boson peak and Ioffe--Regel limits. With increasing disorder, the boson peak frequency $\omega_{BP}$, transverse Ioffe--Regel frequency $\omega_{IR}^T$, and longitudinal Ioffe--Regel frequency $\omega_{IR}^L$ all decrease. We find that there are two ways for the boson peak to form: developing from and coexisting with (but remaining independent of) the transverse van Hove singularity without and with local coordination number fluctuation. In the presence of a single type of disorder, $\omega_{IR}^T\ge \omega_{BP}$, and $\omega_{IR}^T\approx \omega_{BP}$ only when the disorder is sufficiently strong and causes spatial fluctuation of the local coordination number. Moreover, if there is no positional disorder, $\omega_{IR}^T\approx \omega_{IR}^L$. Therefore, the argument that the boson peak is equivalent to the transverse Ioffe--Regel limit is not general. Our results suggest that both local coordination number and positional disorder are necessary for the argument to hold, which is actually the case for most disordered solids such as marginally jammed solids and structural glasses. We further combine two types of disorders to cause disorder in both the local coordination number and lattice site position. The density of vibrational states of the resulting networks resembles that of marginally jammed solids well. However, the relation between the boson peak and the transverse Ioffe--Regel limit is still indefinite and condition-dependent. Therefore, the interplay between different types of disorders is complicated, and more in-depth studies are required to sort it out.

\end{abstract}

\keywords{$|$ disorder $|$ boson peak $|$ Ioffe--Regel limit $|$ amorphous solid}

\pacs{}

\maketitle

\section{Introduction}
\label{sec:intro}

For normal solids such as crystals, the normal modes of vibration are propagating plane waves, i.e., phonons \cite{kittel, ascroft}. Low-frequency phonons form a Debye-like density of states (DOS), $D(\omega)\sim \omega^{d-1}$, where $d$ is the spatial dimension, before reaching the van Hove singularities (two kinks or sharp peaks in the intermediate- and high-frequency regimes corresponding to the Brillouin zone). When a disorder is introduced, the propagating feature of the modes is destroyed above a critical frequency called the Ioffe--Regel frequency, $\omega_{IR}$. When $\omega\approx \omega_{IR}$, the Ioffe--Regel limit is reached, and the wavelength of the modes is comparable to the phonon mean free path, so the modes at $\omega>\omega_{IR}$ lose their particulate nature \cite{ioffe}. Typically, $\omega_{IR}$ decreases with increasing disorder, so it is an important quantity for characterizing the vibrational properties of disordered solids.

In addition to the Ioffe--Regel characteristics, the boson peak is another important vibrational feature of typical disordered solids such as glasses \cite{nakayama,nakayama1,duval,keyes,schirmacher,kantelhardt,grigera,gurevich,sokolov,wuttke,lunkenheimer}. When the DOS is divided by the Debye scaling, the reduced DOS, $D(\omega)/\omega^{d-1}$, exhibits a low-frequency peak at $\omega_{BP}$, which is much lower than the frequency at which the transverse van Hove singularity occurs, implying aggregation of excess soft modes. This boson peak is thought to be the key to understanding abnormal properties of disordered solids, e.g., the glass transition and unusual low-temperature thermal properties and energy transport in glasses \cite{phillipsbook,xu,wyart}. The origin of the boson peak, which has been a topic of vigorous debate, seems to be system- or model-dependent and has been related to a variety of elements, e.g., the transverse Ioffe--Regel limit \cite{shintani,beltukov}, locally favored structures \cite{tanaka1}, local vibration or coherent motion of particle clusters \cite{duval1,angell1}, the phonon-saddle transition (in the energy landscape) \cite{grigera}, extended soft modes arising from marginal stability \cite{silbert,wyart,degiuli}, spatial fluctuation of elastic moduli \cite{schirmacher1,ferrante}, breakdown of the continuum approximation on the mesoscopic length scale \cite{Fl,monaco}, topologically diverse defects of disordered solids \cite{angell2}, and two-level systems \cite{parashin}. Although no conclusion has been reached regarding whether the boson peak has a universal origin, it has been shown that the boson peak is strongly coupled to the instability of disordered solids \cite{wang,singh}. Therefore, clarifying the role of disorder and its effect on the mechanical stability is essential to understanding the boson peak.

To tackle this issue, a straightforward protocol is to observe the evolution of the system with increasing disorder by starting with a perfect crystal. Previous studies have suggested that when force constant disorder is introduced into a lattice, the boson peak evolves from the transverse van Hove singularity of crystals \cite{chumakov2,schirmacher1,taraskin}. In contrast, a recent study revealed another scenario in which, as the particle size polydispersity is continuously increased, the boson peak emerges at the transition from crystals to disordered crystals, where the structure is highly ordered, and the van Hove singularity is still rather pronounced \cite{huatong}. Therefore, evolution from the van Hove singularity may be only a conditional pathway to boson peak formation. The fundamental difference between the two approaches is the type of disorder introduced. It is thus necessary to have a complete picture of how different types of disorder affect the vibration of disordered solids, which is apparently lacking.

Compared with the physical meaning of the boson peak frequency $\omega_{BP}$, that of the Ioffe--Regel frequency $\omega_{IR}$ is more definite. Some studies have demonstrated a link between the boson peak and the transverse Ioffe--Regel limit \cite{shintani,beltukov}. However, it remains unsettled whether this link can be generalized to all types of disordered systems and whether we can claim for sure that the boson peak is equivalent to the transverse Ioffe--Regel limit.

Bearing in mind the issues mentioned above, we study the evolution of two-dimensional mass-spring networks with four types of disorders: disorders in the force constant, lattice site position, local coordination number, and vacancies. These disorders should cover all possibilities. To study them separately will give us a complete and clear picture of the role of each type of disorder in boson peak formation and the link between the boson peak and the transverse Ioffe--Regel limit.

Starting with a perfect triangular lattice, we find that the boson peak emerges and evolves with increasing disorder for all types of disorder. For force constant or lattice site position disorder, which maintain a spatially uniform local coordination number ($z=6$), the boson peak seems to develop from the van Hove singularity. However, the transverse Ioffe--Regel frequency $\omega_{IR}^T$ is always lager than $\omega_{BP}$. For local coordination number or vacancy disorder, the boson peak coexists with residues of the van Hove singularities. $\omega_{IR}^T$ is greater than $\omega_{BP}$ when the disorder is weak and approaches $\omega_{BP}$ only when the disorder is sufficiently strong. Interestingly, as long as the lattice site structure remains perfect, i.e., for force constant or local coordination number disorder, the longitudinal Ioffe--Regel frequency $\omega_{IR}^L$ is always equal to $\omega_{IR}^T$.

We thus obtain several important findings and deductions: (i) the boson peak might evolve from the transverse van Hove singularity only in the absence of local coordination number fluctuation resulting from the removal of springs or lattice sites, (ii) the transverse and longitudinal Ioffe--Regel frequencies depart from each other only in the presence of positional disorder caused by displacing or removing lattice sites, and (iii) the boson peak could be equivalent to the transverse Ioffe--Regel limit only in the presence of both local coordination number fluctuation and positional disorder.

Many disordered solids do possess both local coordination number fluctuation and positional disorder. The joint effects of multiple types of disorder may not be the simple superposition of their individual effects. We thus combine two different types of disorder and compare the vibrational properties of the resulting networks with those of marginally jammed solids.

A packing of particles interacting via contact repulsion undergoes a jamming transition and becomes a jammed solid when the packing fraction is above a critical value $\phi_j$ \cite{liu1,liu2,hecke1,xu1,ohern1,parisi1,torquato,matthieureview}. Jammed solids are disordered in both particle position and local coordination number. One of the most remarkable features of the jamming transition is its isostaticity; i.e., the average coordination number per particle $z$ is equal to $z_c=2d$, which is the minimum requirement for mechanical stability. The vibrational properties of jammed solids are found to be determined by the excess coordination number $\delta z = z- z_c$, especially the formation of a low-frequency plateau in the DOS whose onset frequency is determined by $\delta z$ \cite{silbert,wyart1}.

By randomly removing springs or lattice sites, we obtain an average coordination number comparable to that of the reference jammed solids. When lattice site position disorder is also present, the resulting networks show a DOS similar to that of jammed solids in the entire spectrum, which no single type of disorder can achieve. Our results indicate that local coordination number fluctuation and the resulting small $\delta z$ play the key role in the low-frequency flattening of the DOS, whereas positional disorder results in elimination of the residue of the transverse van Hove singularity. However, when disorders are randomly introduced, the relation between the boson peak and the transverse Ioffe--Regel limit seems quite conditional. Therefore, it is useful to study the effects of a single type of disorder, but more in-depth studies are necessary to understand the complicated interplay among different types of disorder.

\section{Methods}
\label{sec:method}

We start with a perfect two-dimensional triangular lattice with $N$ sites and a lattice constant $a$, and introduce into the system different types of disorder, as defined below. The aspect ratio of the system is $L_x:L_y = 2: \sqrt{3}$. Periodic boundary conditions are applied in both directions. A mass $m$ sits at a lattice site and is connected to its nearest neighbors by a harmonic spring. The interaction potential between site $i$ and its nearest neighbor $j$ is
\begin{equation}
U_{ij} = \frac{1}{2}k_{ij}(r_{ij} - r_{ij}^0)^2,
\end{equation}
where $k_{ij}$ is the force constant; $r_{ij}$ is the separation between $i$ and $j$, i.e., the length of the spring connecting them; and $r_{ij}^0$ is the length of the relaxed spring. For the perfect lattice, $r_{ij}^0=a$, and $k_{ij}=k_0$. Here, we show mainly the results for $N=4096$ systems. The length, energy, and mass are in units of $a$, $k_0a^2$, and $m$, respectively. The frequency is in units of $\sqrt{k_0/m}$.

\begin{figure}
	\subfigure{\includegraphics[width=0.23\textwidth]{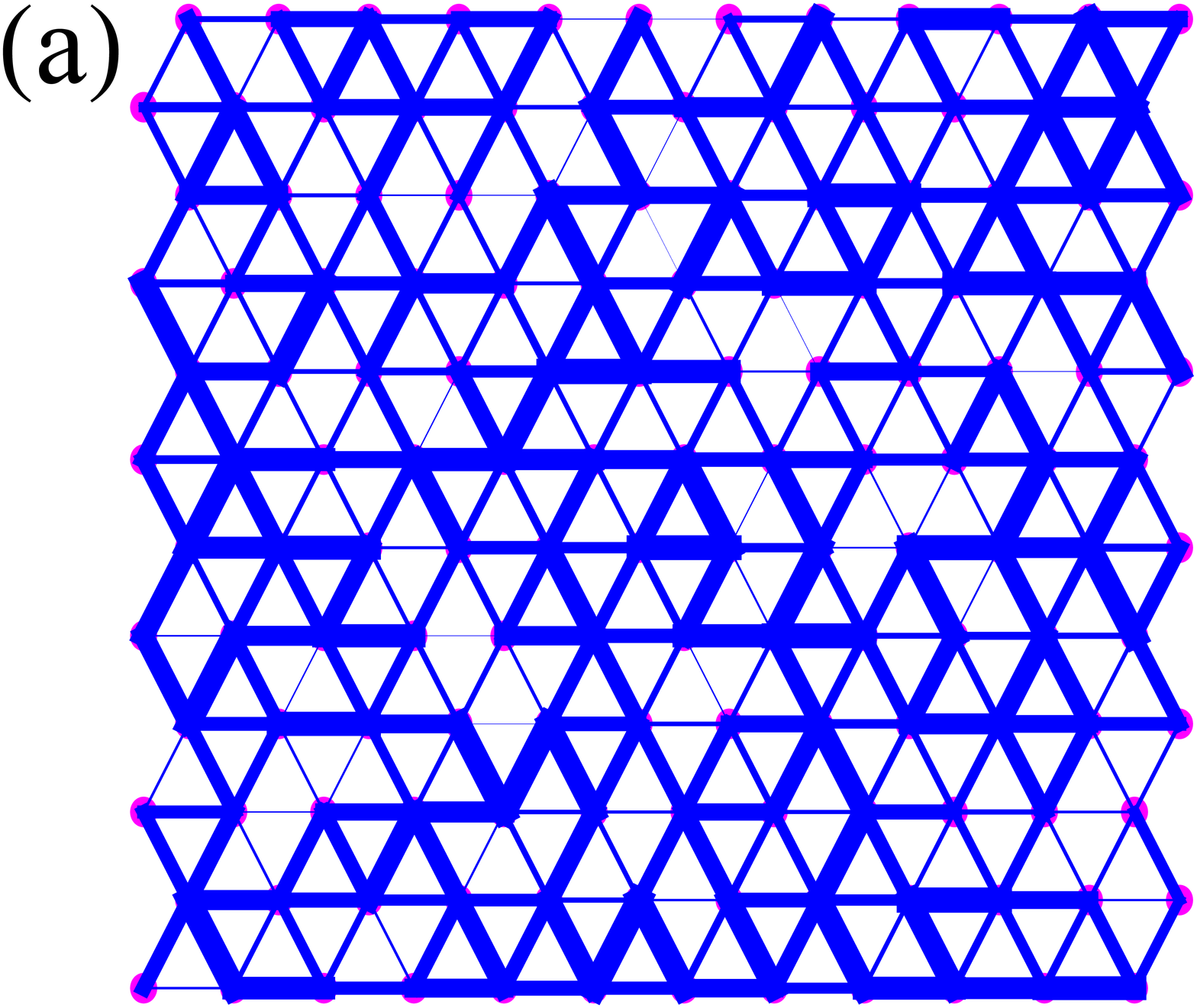}}
	\subfigure{\includegraphics[width=0.23\textwidth]{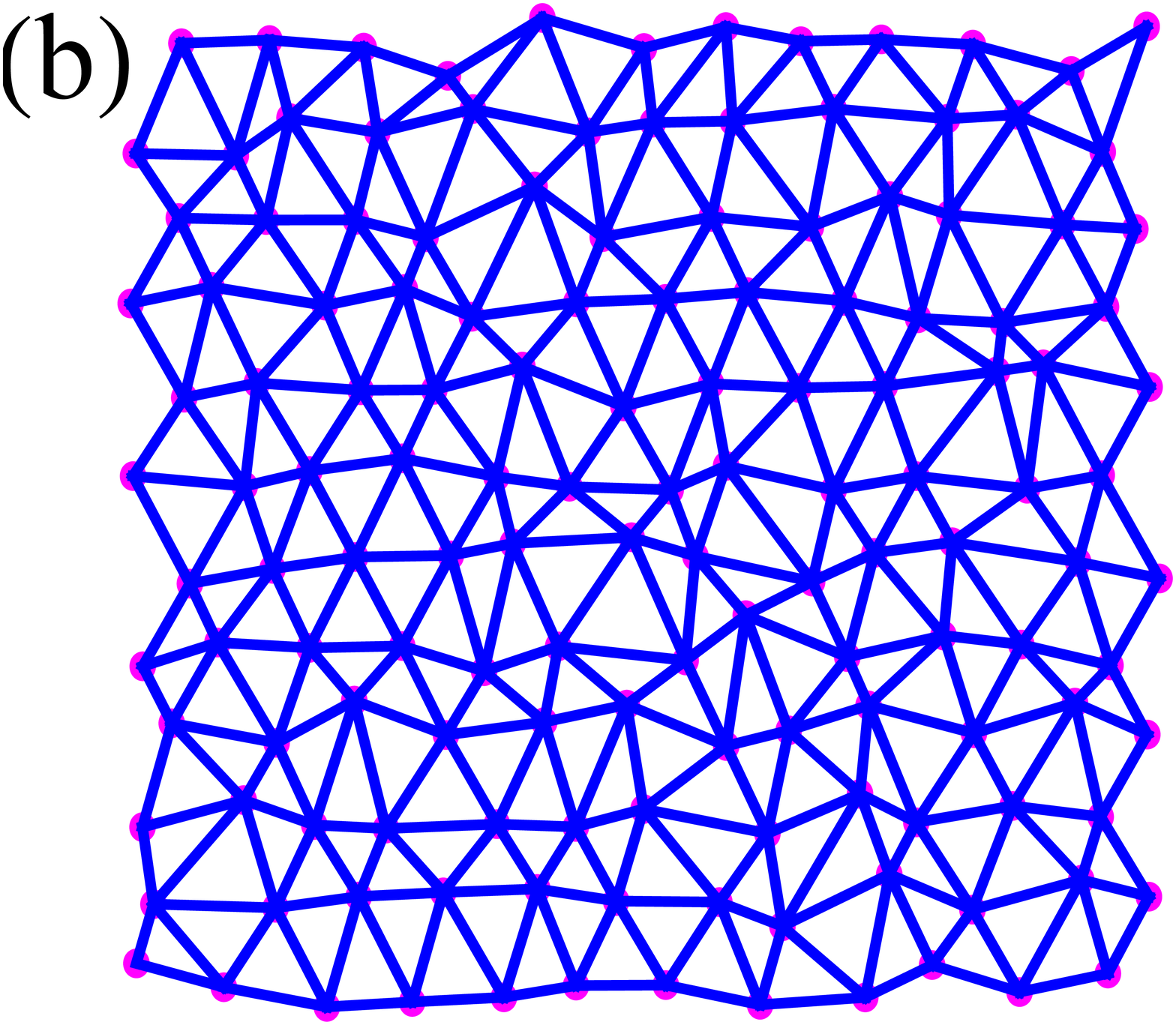}}
	\subfigure{\includegraphics[width=0.23\textwidth]{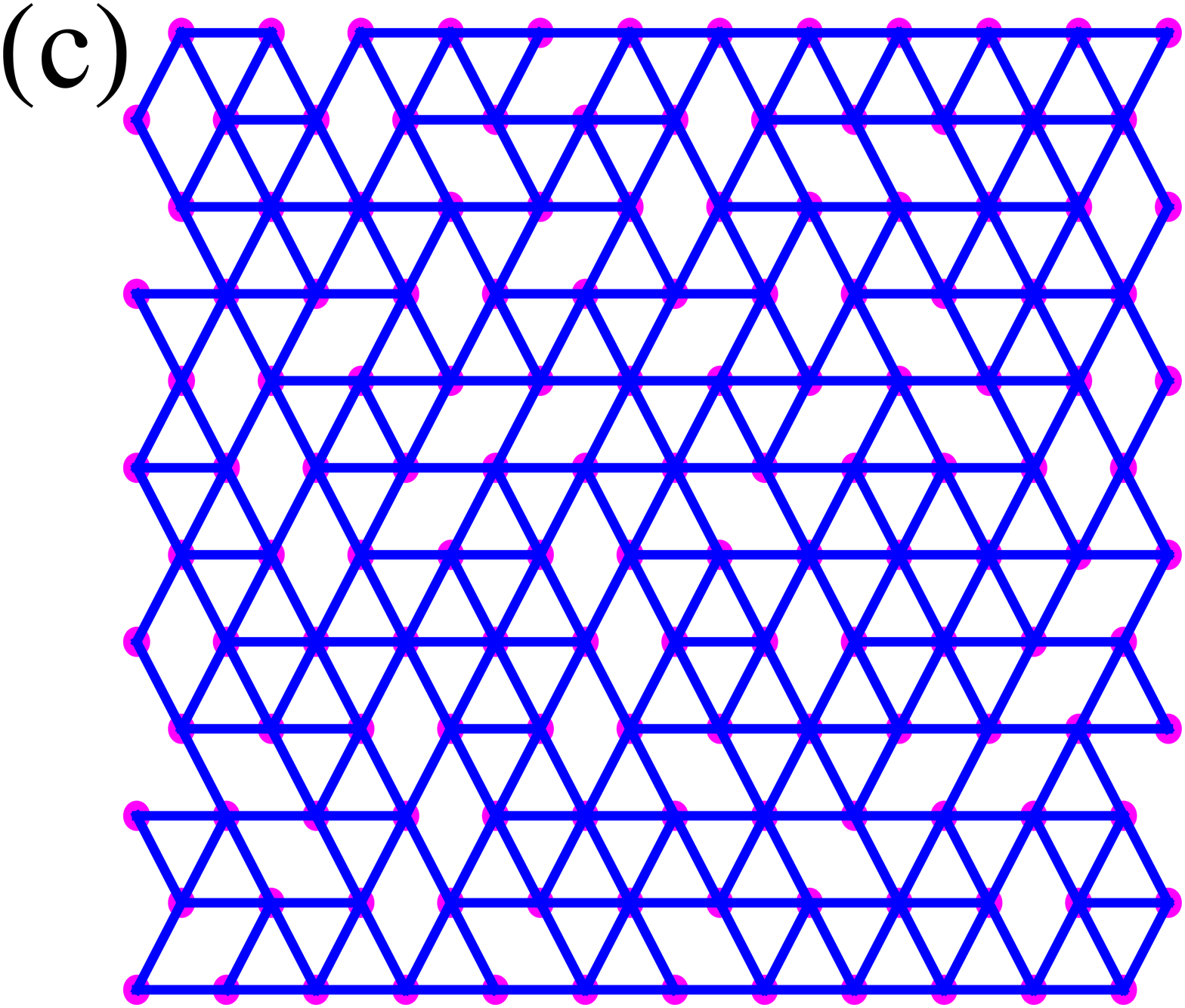}}
	\subfigure{\includegraphics[width=0.23\textwidth]{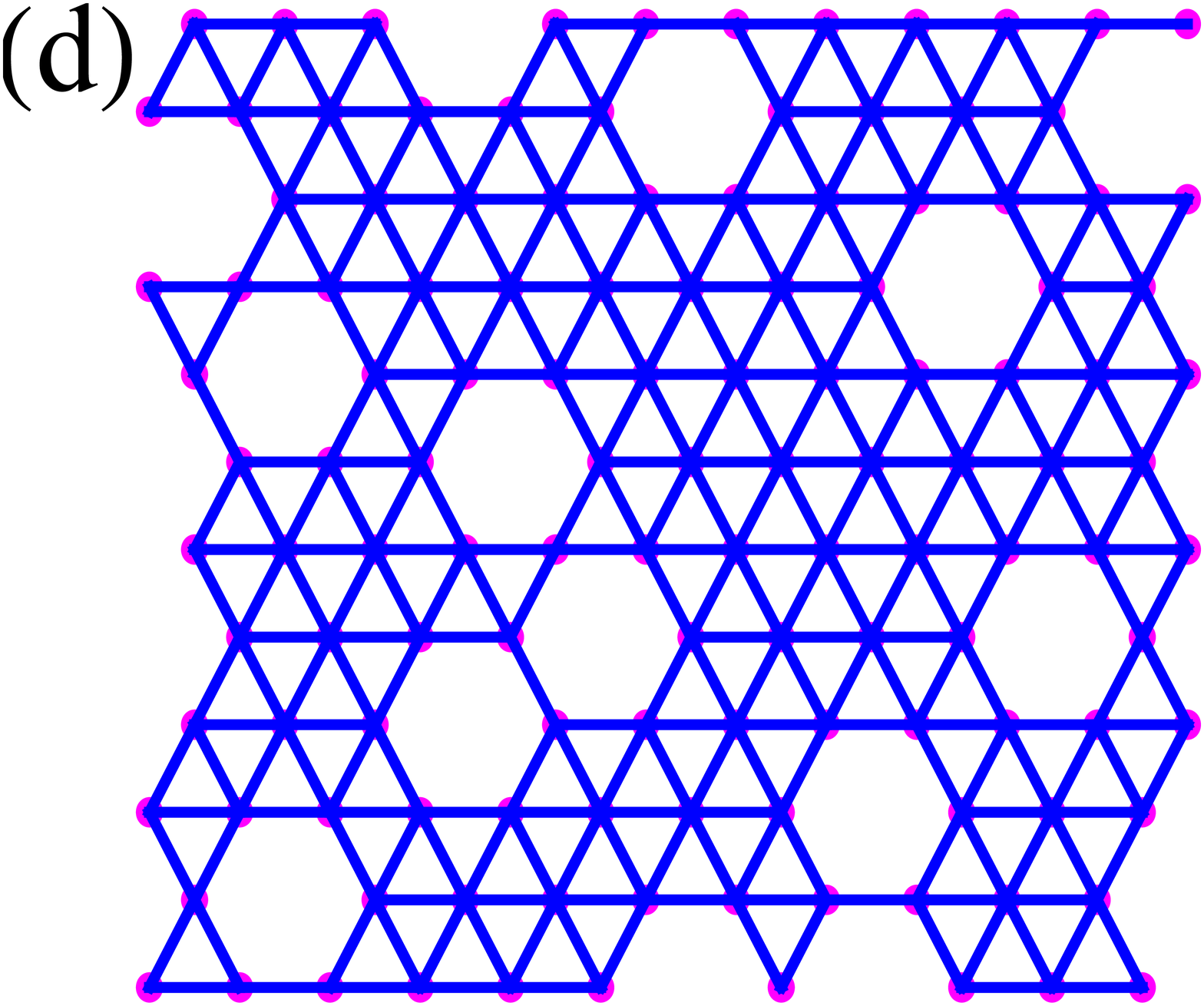}}
\caption{\label{fig:fig1} Illustrations of mass-spring networks with a single type of disorder. Solid circles are lattice sites (masses), and lines are springs connecting neighboring lattices. (a) Force constant disorder, where the thickness of the lines is proportional to the force constant. (b) Lattice site position disorder, where the locations of sites are randomly moved. (c) Local coordination number disorder, where springs are randomly removed. (d) Vacancy disorder, where lattice sites are randomly removed.
}
\end{figure}

We can break the perfection of the lattice and introduce disorder in multiple ways. Figure~\ref{fig:fig1} shows four types of fundamental disorder elements that one can imagine: (a) force constant, $k_{ij}$; (b) lattice site position $\vec{u}_i=\vec{r}_i-\vec{r}_i^p$, where $\vec{r}_i$ and $\vec{r}_i^p$ are the locations of site $i$ for distorted and perfect lattices, respectively; (c) local coordination number [where bonds (springs) are randomly cut]; and (d) vacancy [where sites (masses) are randomly removed]. The types of disorder in (a) and (b) maintain a spatially uniform local coordination number where all sites are always connected by six springs, whereas in (c) and (d), the local coordination number fluctuates. In this work, we specifically define local coordination number disorder as the disorder in (c). The disorder in (d) can be treated as a special case of (c) and is called vacancy disorder to distinguish it from the specified local coordination number disorder. The types of disorder in (a) and (c) maintain perfect lattice site structure, whereas in (b) and (d), positional order is destroyed by site displacements or vacancies.

There is another type of disorder that is not considered here: mass disorder. This type of disorder should affect the properties of normal modes, especially in the presence of extremely heavy or pinned particles \cite{tong_pre}. However, in real systems, particle masses do not vary greatly, and in most model systems, particles are treated as having equal mass. Our preliminary results indicate that the vibrational properties of mass-spring networks are not very sensitive to mass disorder. Therefore, in this work, we concentrate only on the effects of positional and topological disorder.

The networks studied here are unstressed. Namely, all the springs are relaxed, so $r_{ij}^0=a$ for types (a), (c), and (d), whereas it fluctuates for type (b). We obtain all the normal modes of vibration by diagonalizing the Hessian matrix using ARPACK \cite{arpack}. The DOS is calculated from $D(\omega)=\left< \sum_n \delta(\omega - \omega_n)\right> / dN$, where $\omega_n$ is the frequency of the $n$th normal mode of vibration, $\left< .\right>$ denotes the average over tens of independent configurations, and the sum is over all modes. We ensure that the networks considered here are stable by verifying that there are no negative eigenvalues of the Hessian matrix.

To obtain the Ioffe--Regel frequencies, we calculate the dynamical structure factors \cite{shintani,xipeng}
\begin{equation}
	S_{\lambda}(q,\omega)=\frac{q^2}{m\omega^2}\sum_{n}F_{n,\lambda}\delta\left({\omega-\omega_{n}}\right), \label{s1}
\end{equation}
where the sum is over all modes and $\lambda$ denotes T (transverse) or L (longitudinal). Further,
\begin{equation}
	F_{n,L}=|\sum_{j}\left(\vec{e}_{n,j}\cdot\widehat{q}\right){\rm exp}\left(i\vec{q}\cdot\vec{r}_{j}\right)|^2,\label{fl}
\end{equation}
\begin{equation}
	F_{n,T}=|\sum_{j}\left(\vec{e}_{n,j}\times\widehat{q}\right){\rm exp}\left(i\vec{q}\cdot\vec{r}_{j}\right)|^2,\label{ft}
\end{equation}
where the sums are over all sites, $\vec{e}_{n,j}$ is the polarization vector of site $j$ in mode $n$, and $\widehat{q}=\vec{q}/q$, with $\vec{q}$ being the wave vector satisfying the periodic boundary conditions, and $q=|\vec{q}|$.

\section{Results}
\label{sec:results}

In this section, we first show results for only a single type of disorder, from which we expect to extract some rules. Then we will show results for combinations of two types of disorder and compare them with jammed packings of frictionless disks. This decomposition--combination procedure enables us to sort out the role of different types of disorder in determining the vibrational properties of marginally jammed solids.

\subsection{A. Force constant disorder}
\label{subsec:force}

As shown in Fig.~\ref{fig:fig1}(a), force constant disorder is realized by letting the force constant of the springs vary while maintaining perfect lattice structure. We assign the spring connecting sites $i$ and $j$ a force constant $k_{ij}=k_0(1+\eta_k \xi_{ij})$, where $\xi_{ij}$ is a random number uniformly distributed in $[-0.5,0.5]$, and $\eta_k\in [0, 2]$ sets the strength of the disorder.

\begin{figure}
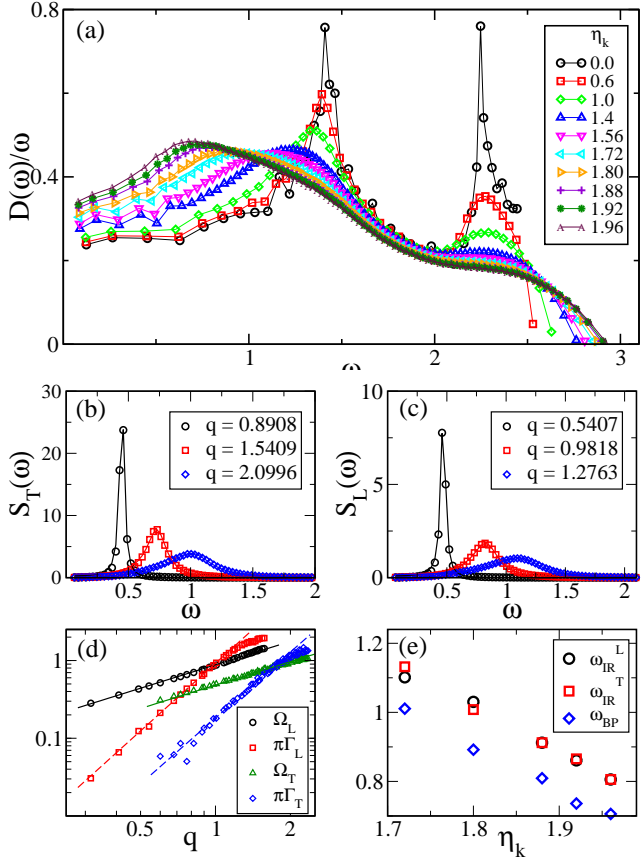

	\includegraphics[width=0.47\textwidth]{Fig2a}
	\includegraphics[width=0.47\textwidth]{Fig2be}	
\caption{\label{fig:fig2} Vibrational properties of networks with force constant disorder $\eta_k$. (a) Evolution of the reduced DOS, $D(\omega)/\omega$, with $\eta_k$. (b) and (c) Examples of the transverse and longitudinal dynamical structure factors (divided by $2N$), $S_{T}(\omega)$ and $S_{L}(\omega)$, respectively, at various $q$ for networks with $\eta_k=1.92$. Lines are fits using Eq.~(\ref{skf}). (d) Dispersion relation $\Omega_{\lambda}(q)$ and excitation broadening $\pi\Gamma(q)$ for $\eta_k=1.92$. Solid and dashed lines have slopes of $1$ and $2$, respectively. (e) Comparison of the boson peak frequency $\omega_{BP}$ and transverse and longitudinal Ioffe--Regel frequencies, $\omega_{IR}^T$ and $\omega_{IR}^L$, respectively, in terms of $\eta_k$.
}
\end{figure}

\begin{figure}
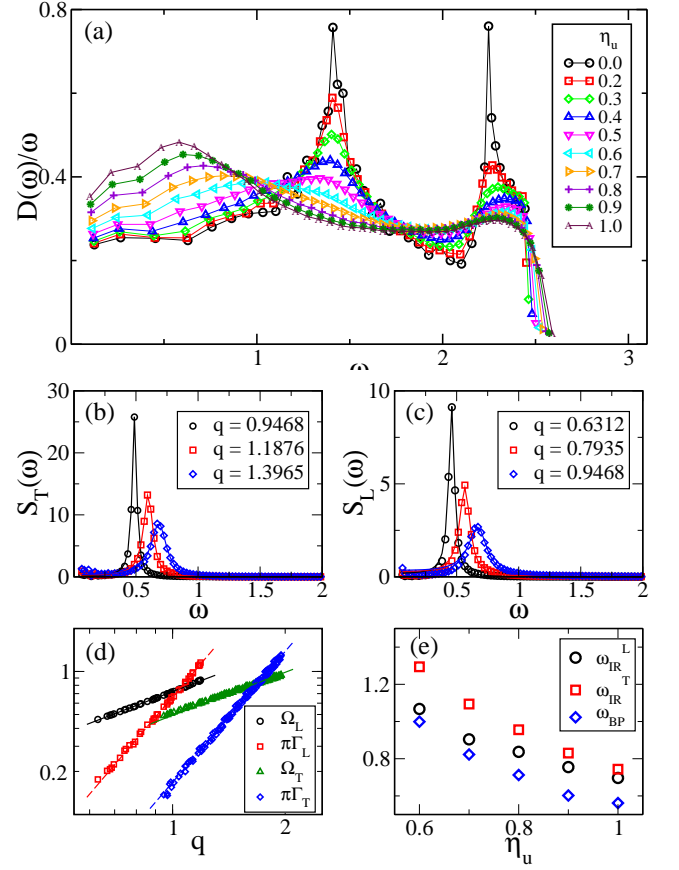

	\includegraphics[width=0.47\textwidth]{Fig3a}
	\includegraphics[width=0.47\textwidth]{Fig3be}
\caption{\label{fig:fig3} Vibrational properties of networks with lattice site position disorder $\eta_u$. (a) Evolution of the reduced DOS, $D(\omega)/\omega$, with $\eta_u$. (b) and (c) Examples of the transverse and longitudinal dynamical structure factors (divided by $2N$), $S_{T}(\omega)$ and $S_{L}(\omega)$, respectively, at various $q$ for networks with $\eta_u=0.9$. Lines are fits using Eq.~(\ref{skf}). (d) Dispersion relation $\Omega_{\lambda}(q)$ and excitation broadening $\pi\Gamma(q)$ for $\eta_u=0.9$. Solid and dashed lines have slopes of $1$ and $2$, respectively. (e) Comparison of the boson peak frequency $\omega_{BP}$ and transverse and longitudinal Ioffe--Regel frequencies, $\omega_{IR}^T$ and $\omega_{IR}^L$, respectively, in terms of $\eta_u$.
}
\end{figure}

Figure~\ref{fig:fig2}(a) shows the evolution of the reduced DOS, $D(\omega)/\omega$, with $\eta_k$. With increasing $\eta_k$, both the transverse and longitudinal van Hove singularities evolve to a broad peak. Further, the transverse van Hove singularity moves to lower frequencies and becomes the low-frequency boson peak. The height of the peak decreases initially and increases after approximately $\eta_k>1.6$. For this case, it is natural to deduce that the boson peak develops from the transverse van Hove singularity, as proposed in Refs. \cite{chumakov2,schirmacher1,taraskin}.

In Figs.~\ref{fig:fig2}(b) and \ref{fig:fig2}(c), we plot examples of the transverse and longitudinal dynamical structure factors using Eqs.~(\ref{s1})--(\ref{ft}). All the curves can be fitted well with \cite{xipeng,hansen}
\begin{equation}
	\begin{split}
S_{\lambda}(q,\omega)
				    & =S_{\lambda, B}(q,\omega)+S_{\lambda,R}(q,\omega)  \\
					& =\frac{A_{\lambda}(q)}{[\omega^2-\Omega_{\lambda}^2{(q)}]^2+\omega^2\Gamma_{\lambda}^2(q)}+\frac{B_{\lambda}(q)}{\omega^2+(D_{\lambda}q^2)^2}, \label{skf}
	\end{split}
\end{equation}
where $S_{\lambda,B}(q,\omega)$ and $S_{\lambda, R}(q,\omega)$ are the Brillouin and Rayleigh components, respectively; $A_{\lambda}(q,\omega)$ and $B_{\lambda}(q, \omega)$ are fitting parameters; $\Omega_{\lambda}(q)$ gives the dispersion relation; $\Gamma_{\lambda}(q)$ characterizes the excitation broadening; and $D_{\lambda}$ is the thermal diffusivity. Figure~\ref{fig:fig2}(d) shows that the dispersion relations for both transverse and longitudinal excitation are linear: $\Omega_{\lambda}=c_{\lambda} q$, where $c_{\lambda}$ is the speed of sound, and $\Gamma_{\lambda}\sim q^2$. The intersection between $\Omega_{\lambda}(q)$ and $\pi\Gamma_{\lambda}(q)$ at $q=q_{IR}$ determines the Ioffe--Regel frequencies \cite{shintani}
\begin{equation}
\omega_{IR}^{\lambda}=\Omega_{\lambda}(q_{IR})=\pi\Gamma_{\lambda}(q_{IR}). \label{ir}
\end{equation}

In Fig.~\ref{fig:fig2}(e), we compare three characteristic frequencies, $\omega_{BP}$ read from Fig.~\ref{fig:fig2}(a), $\omega_{IR}^T$, and $\omega_{IR}^L$. In the entire range of $\eta_k$ studied here, all the frequencies decrease with increasing $\eta_k$, and $\omega_{IR}^T\approx \omega_{IR}^L>\omega_{BP}$. Therefore, in the presence of only force constant disorder, $\omega_{IR}^T$ and $\omega_{IR}^L$ couple to each other, and the gap between $\omega_{IR}^T$ and $\omega_{BP}$ shows no tendency to vanish.

\subsection{B. Lattice site position disorder}
\label{subsec:position}

As shown in Fig.~\ref{fig:fig1}(b), we introduce lattice site position disorder by randomly displacing site $i$ from $\vec{r}_i^p$ to $\vec{r}_i=\vec{r}_i^p+\vec{u}_i$, where $\vec{u}_i=\eta_u \chi_i[{\rm cos}(\theta_i)\hat{x}+{\rm sin}(\theta_i)\hat{y}]$, with $\chi_i\in [-0.5, 0.5]$ and $\theta_i\in[-\pi/2, \pi/2]$ being uniformly distributed random numbers, and $\eta_u$ setting the strength of the disorder. To avoid frequent cross-linking of the springs, we allow $\eta_u$ only in the range from $0$ to $1$. The resulting networks have structural disorder but maintain a spatially uniform local coordination number, $z=6$.

Figure~\ref{fig:fig3}(a) demonstrates that the evolution of $D(\omega)/\omega$ with $\eta_u$ is similar to that for force constant disorder. The transverse van Hove singularity moves to lower frequencies with increasing $\eta_u$, and its height reaches the minimum when $\eta_u\approx 0.6$. Therefore, for this case, the boson peak also seems to develop from the transverse van Hove singularity.

Figures~\ref{fig:fig3}(b) and \ref{fig:fig3}(c) show that both the transverse and longitudinal dynamical structure factors can be fitted well with Eq.~(\ref{skf}), from which we extract $\Omega_{\lambda}(q)$ and $\Gamma_{\lambda}$ and determine the Ioffe--Regel frequency $\omega_{IR}^{\lambda}$ from Eq.~(\ref{ir}) [see examples in Fig.~\ref{fig:fig3}(d)].

The boson peak frequency and two Ioffe--Regel frequencies decrease with increasing $\eta_u$. As has been seen for force constant disorder, $\omega_{BP}$ is also smaller than $\omega_{IR}^T$, and they do not tend to approach each other. However, in contrast to the case of force constant disorder, $\omega_{IR}^T$ is not equal to $\omega_{IR}^L$. Bizarrely, unlike the normal observation of $\omega_{IR}^L>\omega_{IR}^T$ in disordered solids \cite{shintani,xipeng,beltukov}, $\omega_{IR}^L<\omega_{IR}^T$ here. This unusual behavior may result in extraordinary properties of this type of disordered network, which will be discussed elsewhere \cite{junliu}.

\begin{figure}
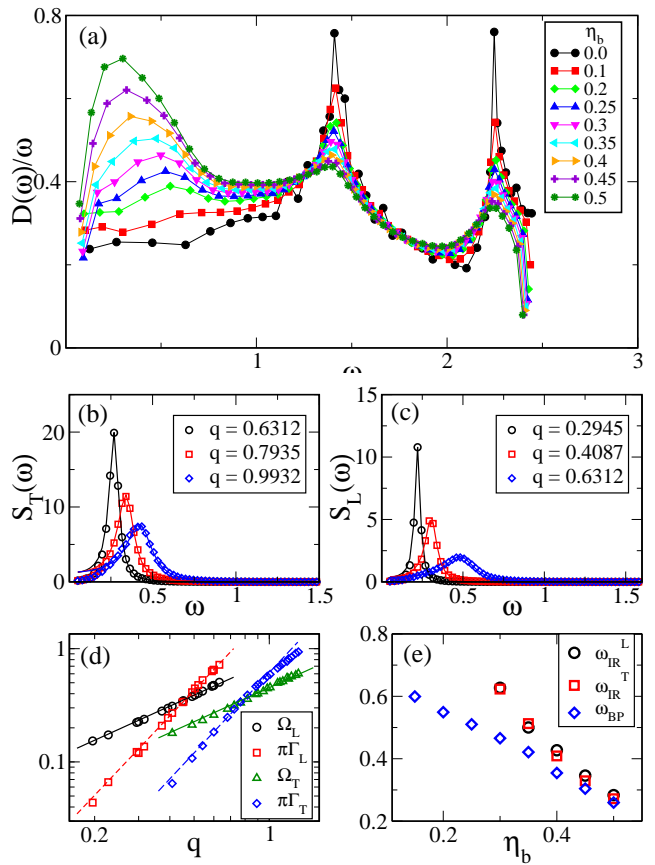

	\includegraphics[width=0.47\textwidth]{Fig4a}
	\includegraphics[width=0.47\textwidth]{Fig4be}
\caption{\label{fig:fig4} Vibrational properties of networks with local coordination number disorder $\eta_b$. (a) Evolution of the reduced DOS, $D(\omega)/\omega$, with $\eta_b$. (b) and (c) Examples of the transverse and longitudinal dynamical structure factors (divided by $2N$), $S_{T}(\omega)$ and $S_{L}(\omega)$, respectively, at various $q$ for networks with $\eta_b=0.45$. Lines are fits using Eq.~(\ref{skf}). (d) Dispersion relation $\Omega_{\lambda}(q)$ and excitation broadening $\pi\Gamma(q)$ for $\eta_b=0.45$. Solid and dashed lines have slopes of $1$ and $2$, respectively. (e) Comparison of the boson peak frequency $\omega_{BP}$ and transverse and longitudinal Ioffe--Regel frequencies, $\omega_{IR}^T$ and $\omega_{IR}^L$, respectively, in terms of $\eta_b$.
}
\end{figure}

\subsection{C. Local coordination number disorder}
\label{subsec:bond}

Local coordination number disorder is introduced by randomly removing $\eta_b N$ springs, as illustrated in Fig.~\ref{fig:fig1}(c). The resulting networks have an average coordination number of $z=6 - 2\eta_b$ and maintain perfect positional order. To maintain mechanical stability, we do not allow $z<4$ or any site to attach to less than three springs. Therefore, $\eta_b$ has values in $[0, 1]$.

Figure~\ref{fig:fig4}(a) indicates that, unlike the case for force constant and lattice site position disorder, the boson peak does not develop from the transverse van Hove singularity. With increasing $\eta_b$, a low-frequency boson peak emerges, grows, and moves to lower frequencies in the presence of two other peaks. These two peaks apparently develop from the van Hove singularities and always stay at roughly the same frequencies. It has been shown that the emergence of the boson peak in this case is a signature of the transition from crystals to disordered crystals, i.e., solids with fairly high crystalline order but sufficiently strong local coordination number fluctuation and the mechanical and vibrational properties of disordered solids \cite{huatong}.

Figure~\ref{fig:fig4}(e) shows that the $\omega_{IR}^T$ and $\omega_{IR}^L$ values obtained from Figs.~\ref{fig:fig4}(b)--(d) are always equal. With increasing $\eta_b$, the Ioffe--Regel frequencies decrease and approach $\omega_{BP}$. All three frequencies agree at sufficiently large $\eta_b$.

\begin{figure}
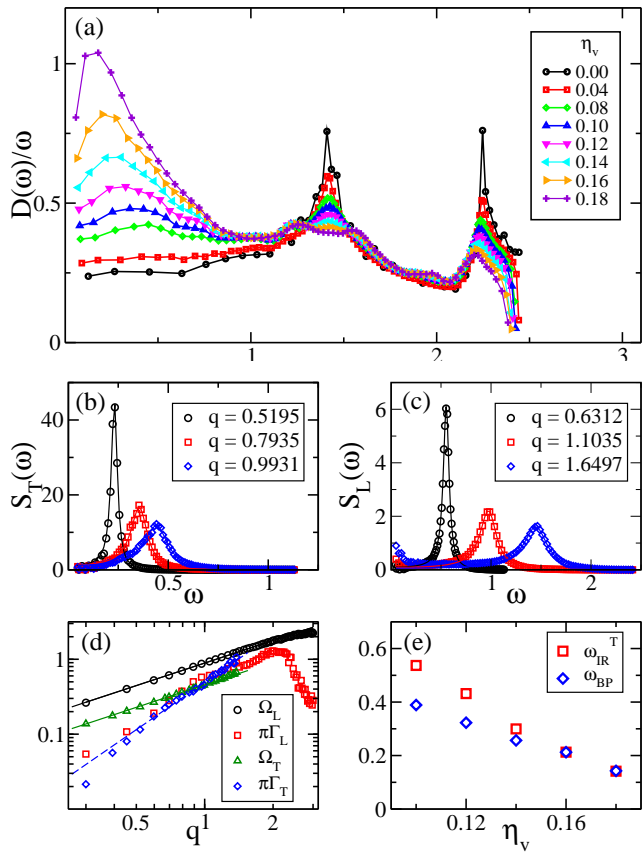

	\includegraphics[width=0.47\textwidth]{Fig5a}
	\includegraphics[width=0.47\textwidth]{Fig5be}
\caption{\label{fig:fig5} Vibrational properties of networks with vacancy disorder $\eta_v$. (a) Evolution of the reduced DOS, $D(\omega)/\omega$, with $\eta_v$. (b) and (c) Examples of the transverse and longitudinal dynamical structure factors (divided by $2N$), $S_{T}(\omega)$ and $S_{L}(\omega)$, respectively, at various $q$ for networks with $\eta_v=0.12$. Lines are fits using Eq.~(\ref{skf}). (d) Dispersion relation $\Omega_{\lambda}(q)$ and excitation broadening $\pi\Gamma(q)$ for $\eta_v=0.12$. Solid and dashed lines have slopes of $1$ and $2$, respectively. (e) Comparison of the boson peak frequency $\omega_{BP}$ and transverse Ioffe--Regel frequency $\omega_{IR}^T$ in terms of $\eta_k$. In the range of $\eta_v$ shown here, the longitudinal Ioffe--Regel frequency $\omega_{IR}^L$ is not well-defined.
}
\end{figure}

\subsection{D. Vacancy disorder}
\label{subsec:vacancy}

In this case, we randomly remove $\eta_v N$ sites and the springs attached to them to realize disorder induced by vacancies, as shown in Fig.~\ref{fig:fig1}(d). To avoid aggregation of vacancies, we deliberately prohibit simultaneous removal of two neighboring sites. The resulting networks deviate from perfect lattice structure owing to the presence of vacancies and have the average coordination number $z=6(1 - 2\eta_v) / (1 - \eta_v)$. To guarantee mechanical stability, $\eta_v$ ranges from $0$ to $0.25$. As mentioned above, this type of disorder also leads to spatial fluctuation of the local coordination number, which is actually a special case of local coordination number disorder, and its effects on the vibrational properties may be similar to those of random removal of springs as discussed in Sec.~C. It destroys the perfect lattice structure, but in a different way from the lattice site position disorder discussed in Sec.~B.

As expected, the evolution of $D(\omega)/\omega$ with $\eta_v$ shown in Fig.~\ref{fig:fig5}(a) looks quite similar to that in Fig.~\ref{fig:fig4}(a). The boson peak apparently does not develop from the transverse van Hove singularity, because it coexists with the peaks corresponding to the van Hove singularities.

As shown in Fig.~\ref{fig:fig5}(d), $\Omega_T(q)$ and $\Omega_L(q)$ extracted from Figs.~\ref{fig:fig5}(b) and \ref{fig:fig5}(c) are still linear. $\pi\Gamma_T(q)$ scales roughly with $q^2$ and intersects $\Omega_T(q)$ at $\omega_{IR}^T$, except that it decays faster with decreasing $q$ when $q$ is small, implying that vacancy disorder does not cause strong scattering of long-wavelength phonons. Vacancy disorder differs from the other three types of disorder in that there is no well-defined scaling for $\Gamma_L(q)$ and $\Omega_L(q)>\pi\Gamma_L(q)$ over the entire spectral range, even though $\omega_{IR}^T$ already exists. Therefore, $\omega_{IR}^L$, if it exists, would exceed the maximum frequency of all modes, which is not meaningful. Only when $\eta_v$ is sufficiently large do we see the intersection between $\Omega_L(q)$ and $\pi\Gamma_L(q)$ and estimate $\omega_{IR}^L$, which is larger than $\omega_{IR}^T$.

In Fig.~\ref{fig:fig5}(e), we compare only $\omega_{BP}$ with $\omega_{IR}^T$, because the longitudinal excitations do not exhibit an Ioffe--Regel limit in the $\eta_v$ range shown here. Both frequencies decrease with increasing $\eta_v$. $\omega_{IR}^T$ is larger than $\omega_{BP}$ when $\eta_v$ is small and is equal to $\omega_{BP}$ at large $\eta_v$. For this case, we may argue that the boson peak corresponds to the transverse Ioffe--Regel limit when the disorder is sufficiently strong.

\subsection{E. Rules summarized from cases of a single type of disorder}

As discussed in the Methods section, the networks with different types of disorder share some common structural or topological features. Networks with disorder in the force constant or lattice site position maintain a spatially uniform local coordination number ($z=6$), whereas $z$ fluctuates when springs or sites are randomly removed. The lattice site structure remains perfect with disorder in the force constant or local coordination number, whereas positional order is destroyed in the presence of disorder in the lattice site position or vacancies, but in different ways. By considering the structural, topological, and vibrational properties of disordered networks with four types of disorder, we can summarize some rules and enhance our understanding of the origin of the boson peak and its relation to the transverse Ioffe--Regel limit.

First, all types of disorder cause the emergence of the boson peak and Ioffe--Regel limits. With increasing disorder strength, the boson peak and Ioffe--Regel frequencies all decrease.

Our study reveals two possible mechanisms for formation of the boson peak. If there is no local coordination number fluctuation, the boson peak develops from the transverse van Hove singularity, regardless of whether the lattice structure is ordered. In contrast, local coordination number fluctuation leads to the coexistence of the boson peak and residues of the van Hove singularities.

Regarding the relationship between the boson peak and the transverse Ioffe--Regel limit, our results question the generality of the argument that they are equivalent. We would argue that they match only under certain circumstances. $\omega_{BP}\approx \omega_{IR}^T$ only when a sufficiently large number of springs or lattice sites are randomly removed, and the boson peak does not simply develop from the transverse van Hove singularity. Otherwise, $\omega_{IR}^T>\omega_{BP}$. Even though $\omega_{BP}\approx \omega_{IR}^T$ could be true, the transverse nature of the boson peak is questionable when $\omega_{IR}^L\approx \omega_{IR}^T$, which seems to be true as long as there is no positional disorder. In the presence of a single type of disorder, vacancy disorder is the only one showing that the boson peak might agree with the transverse Ioffe--Regel limit.

Note that even though the presence of positional disorder leads to separation of $\omega_{IR}^T$ and $\omega_{IR}^L$, there are two different consequences: $\omega_{IR}^L<\omega_{IR}^T$ for lattice site position disorder, whereas $\omega_{IR}^L>\omega_{IR}^T$ for vacancy disorder. Therefore, although $\omega_{IR}^T$ and $\omega_{IR}^L$ could separate, vacancy disorder, but not lattice site position disorder, seems to be a good choice for boosting the agreement between the boson peak and the transverse Ioffe--Regel limit.

\begin{figure}
	\includegraphics[width=0.4\textwidth]{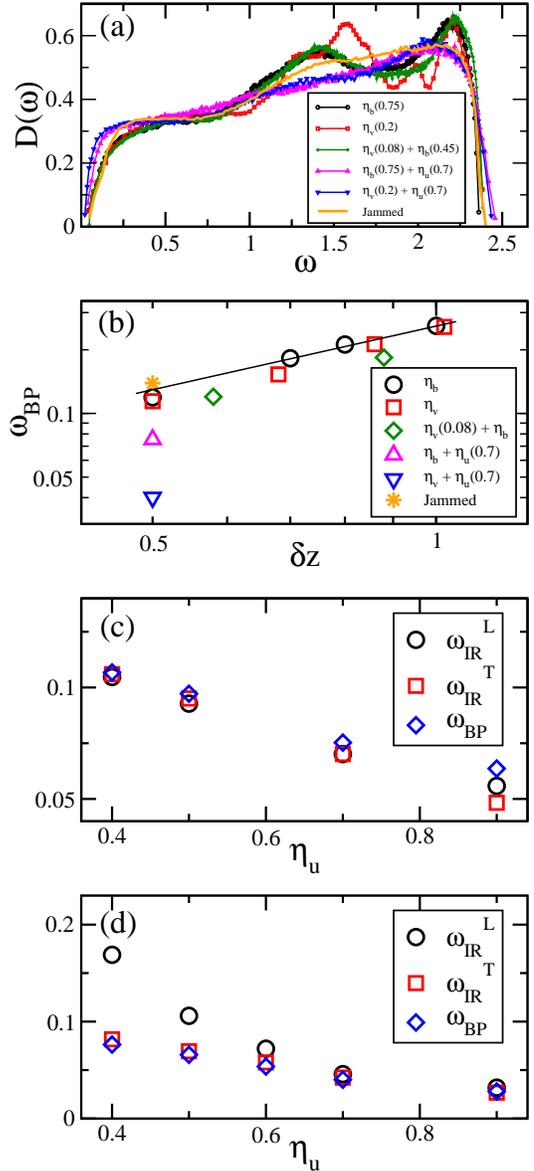}
\caption{\label{fig:fig6} (a) Density of states $D(\omega)$ for disordered networks and unstressed jammed solids, all with an excess average coordination number $\delta z=0.5$. The values of the disorder strength are shown in parentheses in the legend. (b) Boson peak frequency $\omega_{BP}$ versus $\delta z$ for different systems. Line shows $\omega_{BP}\sim \delta z$. (c) and (d) Lattice site position disorder ($\eta_u$) dependence of characteristic frequencies of networks with $\delta z=0.5$ induced by local coordination number disorder and vacancy disorder, respectively.
}
\end{figure}

\subsection{F. Combination of two types of disorder and comparison to jammed solids}

Most typical disordered solids, e.g., marginally jammed solids and structural glasses, contain multiple types of disorder. Although the results presented above help clarify our picture of how a single type of disorder affects the vibration of networks, the effects of competition or cooperation among different types of disorder on the vibrational properties of disordered solids remain elusive. In this section, we show results for networks with two different types of disorder and see how their joint effects agree with or deviate from the expectations based on our observations of a single type of disorder.

Here, we consider only combinations of lattice site position, local coordination number, and vacancy disorder, in order to compare the resulting networks with marginally jammed solids with harmonic repulsion. To obtain jammed solids, we can simply replace lattice site $i$ of the networks studied above with a particle of mass $m$ and diameter $\sigma_i$. Particles $i$ and $j$ interact via the potential
\begin{equation}
U_{ij}=\frac{k_0}{2}\left( \sigma_{ij}-r_{ij} \right)^{2}\Theta\left(1-\frac{r_{ij}}{\sigma_{ij}}\right), \label{eq:hertzian}
\end{equation}
where $\sigma_{ij}$ is the sum of their radii, and $\Theta(x)$ is the Heaviside step function. The force constant of the pair interaction is constant, so we do not need to include force constant disorder in this section. We study a polydisperse system with $\sigma_i=\sigma(1+\eta \chi_i)$, where $\chi_i$ is a random number uniformly distributed in $[-0.5, 0.5]$, and $\eta\in [0, 2]$ sets the particle size polydispersity. We use the fast inertial relaxation engine algorithm~\cite{fire} to minimize the total potential energy of the system, $U=\sum_{ij} U_{ij}$, and obtain mechanically stable jammed states, where the sum is over all pairs of particles. The units of mass, length, and energy are $m$, $\sigma$, and $k_0\sigma^2$, respectively, so the frequency is in units of $\sqrt{k_0/m}$. Here, we show the results for two-dimensional $N=4096$ jammed solids with $\eta=0.4$ at a pressure $p=8.8\times10^{-3}$, which have an average coordination number $z\approx 4.5$. Note that we study disordered networks in which all springs are relaxed. We thus replace the particle interactions of jammed solids with relaxed springs by setting the first derivative of the potential energy to zero while maintaining its second derivative. These jammed solids are thus unstressed, and their pressure is reset to zero. The unstressed solids are more stable because the existence of pressure destabilizes the system \cite{xu}.

It has been proposed that the low-frequency vibration of marginally jammed solids is determined by the excess average coordination number above isostaticity, $\delta z =z-z_c$ \cite{silbert,wyart}. We thus generate disordered networks with $\delta z=0.5$, that of jammed solids, by introducing local coordination number or vacancy disorder. We are concerned about three combinations: (I) local coordination number and lattice site position disorder, (II) vacancy and lattice site position disorder, and (III) local coordination number and vacancy disorder. Jammed solids are disordered in both local coordination number and particle position. Therefore, types (I) and (II) should induce networks with structure and topology more closely resembling those of jammed solids than type (III), because the remaining lattice sites for type (III) are still on the sites of a perfect triangular lattice.

In Fig.~\ref{fig:fig6}(a), we compare the DOSs of different networks with jammed solids. Let us first compare networks with only local coordination number or vacancy disorder ($\eta_b=0.75$ or $\eta_v=0.2$) and the combination of the two ($\eta_b=0.45$ and $\eta_v=0.08$), all with $\delta z =0.5$, the value for jammed solids. Because of isostaticity, there exists a low-frequency plateau in the DOS of jammed solids, whose onset frequency has been shown to be $\omega^* \sim \delta z$ for harmonic repulsion \cite{silbert,wyart1}. The boson peak identified from $D(\omega)/\omega$ normally occurs at $\omega_{BP}<\omega^*$, on the ramp of $D(\omega)$ before the plateau. $\omega_{BP}$, like $\omega^*$, is also proportional to $\delta z$. Therefore, $\omega^*$ is usually treated like $\omega_{BP}$, especially for unstressed jammed solids, in which the low-frequency DOS is steeply ramped and thus $\omega_{BP}\approx \omega^*$. Interestingly, all three $D(\omega)$ curves exhibit a low-frequency plateau, and their low-frequency region almost overlaps that of jammed solids. Figure~\ref{fig:fig6}(b) shows that the three networks have a boson peak frequency similar to that of jammed solids (compare the $\omega_{BP}$ values at $\delta z =0.5$), and $\omega_{BP}\sim \delta z$ is roughly true for different values of $\eta_b$ and $\eta_v$. Therefore, our observations support the argument that the low-frequency vibration of systems with contact repulsion (marginally stable solids) is determined by the excess average coordination number above isostaticity. Consequently, the boson peak does not simply develop from the transverse van Hove singularity.

Although the low-frequency region of the DOS already looks similar to that of jammed solids, as discussed above, the masses of the three types of networks are still located at the original sites of the perfect triangular lattice. This leads to residues of the van Hove singularities, manifested as two peaks in the intermediate- and high-frequency regimes, which deviate from the DOS of jammed solids. Because the transverse van Hove singularity moves to lower frequencies and develops into the boson peak with increasing $\eta_u$, we thus superimpose lattice site position disorder on local coordination number or vacancy disorder. With increasing $\eta_u$, the residue of the transverse van Hove singularity indeed decays and eventually disappears. As shown in Fig.~\ref{fig:fig6}(a), for a sufficiently large $\eta_u$, e.g., $\eta_u=0.7$ here, the DOSs of the resulting networks are more like that of jammed solids in the entire spectrum.

Our study here thus reveals that local coordination number or vacancy disorder and lattice site position disorder are important in determining the vibrational properties of jammed solids in the low- and intermediate-to-high-frequency regimes, respectively. Because all the types of disorder considered in this work induce the boson peak, it is expected that the superposition of two types of disorder may quantitatively affect the boson peak frequency. As shown in Fig.~\ref{fig:fig6}(a), the addition of $\eta_u=0.7$ extends the plateau in $D(\omega)$ to slightly lower frequencies, so the combination of two types of disorder leads to a lower boson peak frequency than those of any single type of disorder.

For the jammed solids studied here, $\omega_{BP}$ is slightly higher than $\omega_{IR}^T$, which we may treat as equal. However, $\omega_{IR}^L$ is apparently higher than $\omega_{BP}$. Therefore, we may claim that the boson peak corresponds to the transverse Ioffe--Regel limit. In Figs.~\ref{fig:fig6}(c) and \ref{fig:fig6}(d), we compare $\omega_{BP}$, $\omega_{IR}^T$, and $\omega_{IR}^L$ for disordered networks with increasing $\eta_u$ at a constant $\delta z=0.5$. All the frequencies decrease with increasing $\eta_u$, as expected. At $\eta_u=0.7$, however, when $D(\omega)$ looks similar to that of jammed solids, all three frequencies are still roughly the same, unlike the case for jammed solids. By randomly introducing two types of disorder, we still cannot reproduce the equivalence between the boson peak and the transverse Ioffe--Regel limit. However, the combination of lattice site position and vacancy disorder may provide some clues. As discussed above, both types of disorder lead to the departure of $\omega_{IR}^L$ from $\omega_{IR}^T$, but vacancy disorder pushes $\omega_{IR}^L$ to much higher frequencies. The local coordination number fluctuation induced by vacancy disorder might be important in jammed solids, so more work is required to define and identify its effects.

\section{Discussion and conclusions}

In this paper, we study the role of different types of disorder in determining the vibrational properties of disordered systems. By separately introducing four types of disorder to a two-dimensional triangular lattice, we obtain findings that enhance our understanding of the origin of the boson peak and the relationship between the boson peak and the transverse Ioffe--Regel limit. The boson peak either develops from or coexists with the transverse van Hove singularity, depending on whether the local coordination number is spatially uniform. The boson peak frequency is equal to the transverse Ioffe--Regel frequency only when a sufficiently large number of springs or lattice sites are randomly removed. The equivalence between the boson peak and the transverse Ioffe--Regel frequency suggested in the literature is thus questionable, which is further questioned by the fact that the transverse and longitudinal Ioffe--Regel frequencies are equal, except in the presence of vacancy disorder.

By combining two types of disorder, we obtain a DOS similar to that of jammed solids with the same excess average coordination number. We thus conclude that the local coordination number fluctuation caused by randomly removing springs or lattice sites is important to determine the low-frequency vibration of marginally jammed solids, whereas lattice site position disorder affects mainly the intermediate- and high-frequency vibration by eliminating the transverse van Hove singularity. The combination of two types of disorder leads to a decrease in the characteristic frequencies. However, random superposition of two types of disorder does not always produce disordered networks exhibiting equivalence between the boson peak and the transverse Ioffe--Regel limit.

The decomposition--combination process applied in this work helps us to sort out the role of different types of disorder and recover some unusual vibrational properties of marginally jammed solids. However, we still see discrepancies between the resulting networks and jammed solids. Note that in this work, we introduce disorders in a random way, as long as the networks are stable. The discrepancies imply that this random way does not capture all of the structural or topological features of jammed solids. To produce networks with vibrational properties more similar to those of jammed solids, we may need to introduce types of disorder selectively. To achieve this goal, we can in turn attempt to decompose different types of disorder in jammed solids. Then, some highly useful techniques, such as machine learning \cite{cubuk}, may play an important role.

\section{Acknowledgement}
This work is supported by National Natural Science Foundation of China Grants No.~21325418 and 11574278, and Fundamental Research Funds for the Central Universities Grant No.~2030020028. We also thank the Supercomputing Center of University of Science and Technology of China for computer times.


\end{document}